\documentclass[twocolumn]{aastex62}

\submitjournal{ApJL}

\shorttitle{Wetter Stratospheres on High Obliquity Planets}
\shortauthors{Kang}

\begin{document}

\title{Wetter Stratospheres on High Obliquity Planets}

\correspondingauthor{Wanying Kang}
\email{wanyingkang@g.harvard.edu}

\author[0000-0002-4615-3702]{Wanying Kang}
\affil{School of Engineering and Applied Sciences\\
  Harvard University\\
  Cambridge, MA 02138, USA}


\begin{abstract}
  We investigate how obliquity affects stratospheric humidity using a 3D general circulation model and find the stratosphere under high obliquity could be over 3 orders of magnitude moister than under the low obliquity equivalent, even with the same global annual mean surface temperature. Three complexities that only exist under high obliquity are found to be causally relevant. 1) Seasonal variation under high obliquity causes extremely high surface temperatures to occur during polar days, moistening the polar air that may eventually enter the stratosphere. 2) Unlike the low obliquity scenario where the cold trap efficiently freezes out water vapor, the high obliquity stratosphere gets most of its moisture input from high latitudes, and thus largely bypasses the cold trap. 3) A high obliquity climate tends to be warmer than its low obliquity equivalent, thus moistening the atmosphere as a whole. We found each of the above factors could significantly increase stratospheric humidity. These results indicate that, for an earth-like exoplanet, it is more likely to detect water from surface evaporation if the planet is under high obliquity. The water escape could cause a high obliquity planet to loss habitability before the runaway greenhouse takes place.
\end{abstract}


\section{Introduction} \label{sec:intro}
Before getting to the catastrophic runaway greenhouse state \citep{Ingersoll-1969:runaway}, water escape could have rendered a planet uninhabitable \citep{Kasting-1988:runaway}. If the Earth had a 3000 ppm water vapor mixing ratio in the stratosphere, all the surface water would have been lost within 1 billion years, significantly reducing the chance to host life. According to 1D models, such a stratospheric water concentration can be achieved at 340K surface temperature, corresponding to 1.1 times present day insolation \citep[assuming a moist adiabatic troposphere merged to an isothermal stratosphere,][]{Kasting-Pollack-1983:loss, Kasting-Whitmire-Reynolds-1993:habitable, Wordsworth-Pierrehumbert-2013:water}. As surface temperatures rise, tropopause moves upward, and the same saturated vapor pressure corresponds to a higher water vapor mixing ratio. 

While 1D models provide an acceptable approximation of the upper atmosphere humidity, the seasonal and spatial variation cannot be explicitly represented. Seasonal variation could make a difference by changing the maximum surface temperature. Under high obliquity, continuous direct radiation during the polar days could give rise to very high surface temperatures that are not possible under low obliquity conditions. This high surface temperature would increase the atmosphere's water capacity, leading to a moister stratosphere. \citet{Spiegel-Menou-Scharf-2009:habitable} has attempted to parameterize the seasonal variation in a 1D energy balance model, and they demonstrated the strong seasonal cycle under high obliquity. In the context of the early Mars climate, the tropospheric water vapor is predicted by GCMs to be significantly more abundant during the high obliquity periods \citep{Jakosky-Henderson-Mellon-1995:chaotic, Mischna-Richardson-Wilson-et-al-2003:orbital}, given an infinite water reservoir at the surface. If this more abundant water vapor could be brought to the stratosphere by atmospheric general circulation, water escape may become much faster.

Another factor ignored in 1D models is the horizontal inhomogeneity. The cold trap (the altitude that water vapor stops condensing\footnote{People sometimes define the cold trap to be location of the minimum temperature in the atmosphere, because no condensation is expected above this level.}) is one type of spatial inhomogeneity that can significantly affect the water vapor concentration in the upper atmosphere. The cold trap temperature controls the final saturated vapor pressure, and its altitude controls the total air pressure. Together they determine the water abundance in the upper atmosphere. \citet{Wordsworth-Pierrehumbert-2013:water} demonstrated that water escape would not be enhanced by increasing CO$_2$, even though the surface temperature is increased. This is because high CO$_2$ also cools the stratosphere, thus making the cold trap more efficient. However when we consider 3 dimensions, it is possible that water vapor enters the stratosphere without going through the cold trap. Understanding stratospheric circulation thus becomes crucial.

On the Earth, the stratospheric circulation, referred to as Brewer-Dobson circulation \citep[BD circulation][]{Brewer-1949:evidence,Dobson-1956:origin}, helps prevent water from escaping. Planetary eddies, generated in the mid-latitude weather systems, propagate upward to the stratosphere, depositing easterly momentum there and driving poleward motion. This motion pumps air upward into the stratosphere at the equator, while making the equatorial lower stratosphere the coldest point on the planet \citep{Holton-Gettelman-2001:horizontal}. Since the upward motion collocates with the cold trap on Earth, water vapor is frozen out before it can get high enough to escape \citep{Holton-Haynes-McIntyre-et-al-1995:stratosphere, Butchart-2014:brewer}. Therefore, knowing the stratospheric circulation is crucial to understanding the distribution of tracers, such as photo-chemical products, aerosols, clouds and water vapor, which could in principle be detected on exoplanets. Knowing the stratospheric circulation is also crucial to understanding the mechanisms of water escape, which of course would have important consequences for habitability. The stratospheric circulation and the implications on water escape has been discussed in the context of tidally-locked exoplanets \citep{Carone-Keppens-Decin-et-al-2017:stratosphere, Fujii-Del-Amundsen-2017:nir}, and Earth-like planets \citep{Leconte-Forget-Charnay-et-al-2013:increased, Popp-Schmidt-Marotzke-2016:transition}, while the high obliquity condition has yet to be well explored. 

High obliquity planets are thought to widely exist in the universe as a result of angular momentum exchange between different orbits in a three-body system \citep[Lidov-Kozai cycle][]{Naoz-2016:eccentric}, planet-planet scattering \citep{Chatterjee-Ford-Rasio-2010:how}, secular resonance-driven spin-orbit coupling \citep{Millholland-Laughlin-2019:obliquity}, and giant impacts. In our solar system, Mars's obliquity chaotically varies from 0 to 60 degree \citep{Laskar-Robutel-1993:chaotic}, and Venus and Uranus have obliquities close to 180 and 90 degree respectively \citep{Carpenter-1966:study}. Obliquity has significant impacts on climate. The high obliquity planets have been found to have less ice coverage and higher surface temperature than on their low obliquity equivalents \citep{Jenkins-2003:gcm, Linsenmeier-Pascale-Lucarini-2015:climate, Kilic-Raible-Stocker-2017:multiple, Kilic-Lunkeit-Raible-et-al-2018:stable, Kang-2019:mechanisms, Williams-Kasting-1997:habitable, Gaidos-2004:seasonality, Spiegel-Menou-Scharf-2009:habitable, Rose-Cronin-Bitz-2017:ice, Armstrong-Barnes-Domagal-Goldman-et-al-2014:effects}, and to have drastically different general circulation and planetary eddy behavior changes \citep{Ferreira-Marshall-OGorman-et-al-2014:climate, Kang-Cai-Tziperman-2019:tropical, Ohno-Zhang-2019:atmospheres}. This raises several interesting questions: 1) Would the high obliquity planets be more vulnerable to water escape due to its greater warmness and its strong seasonal variation? 2) How do the drastically different eddies on high obliquity planets affect stratospheric circulation? 3) Does the cold trap exist on high obliquity planets, and if so, can it block water from entering the stratosphere?

In this work, we use a 3D GCM to investigate how the stratospheric water vapor concentration changes with insolation, under high and low obliquity scenarios. The results would help constrain the inner edge of the habitable zone for both scenarios, help evaluate the detectability of exoplanet surface water from spacecraft observation, and possibly provide a mechanism to explain water escape during the early history of the Mars.

\section{Methods}
\label{sec:methods}
The model used here is Community Earth System Model version 1.2.1 \citep[CESM,][]{Neale-Chen-Gettelman-2010:description}, modified by \citet[][code are available on GitHub\footnote{ https://github.com/storyofthewolf/ExoRT and https://github.com/storyofthewolf/ExoCAM}]{Wolf-Toon-2015:evolution, Wolf-2017:assessing, Wolf-Shields-Kopparapu-et-al-2017:constraints, Kopparapu-Wolf-Arney-et-al-2017:habitable, Haqq-Misra-Wolf-Joshi-et-al-2018:demarcating} to include mainly the following two features: 1) more realistic radiation calculation resulting from increased spectral resolution, updated spectral coefficients based on the HiTran 2012 database \citep{Rothman-Gordon-Babikov-et-al-2013:hitran2012}, and a new continuum opacity model \citep{Paynter-Ramaswamy-2014:investigating}, and 2) more frequent sub-step dynamic adjustment to improve numerical stability. We consider H$_2$O as the only greenhouse gas in the atmosphere for simplicity, while ignoring the CO$_2$ absorption. Thanks to the fine spectral resolution, this radiation scheme was shown to be more robust at the high temperature end, while the default CESM radiative transfer model underestimates both longwave and shortwave water vapor absorption \citep{Yang-Leconte-Wolf-et-al-2016:differences}. This advantage is crucial for the estimation of the inner edge habitable zone under low and high obliquity scenarios.
The atmosphere circulation is simulated by a finite-volume dynamic core, with approximately 1.9 degree horizontal resolution and 40 vertical layers extending to 0.8 mb. This atmosphere model is coupled with a 50 meter deep slab ocean. Horizontal ocean heat transport is not included for simplicity, since it has been shown to play a minor role in the surface temperature, compared to a large change in obliquity \citep{Jenkins-2003:gcm}. Sea ice is simulated using Community Ice CodE (CICE) version 4, which is part of CESM 1.2.1.

We perform two series of experiments forced by a slowly increasing insolation, one at zero obliquity and the other at 80 degree obliquity. We try to cover the whole habitable range, with the lowest insolation corresponding to an almost snowball state and the highest insolation corresponding to an almost run-away greenhouse state. For both experiments, we vary the insolation from 1360 W/m$^2$ to 1750 W/m$^2$ in 100 years. The upper limit, 1750 W/m$^2$, is chosen to be just below the runaway greenhouse threshold. We choose to do a transient simulation rather than a series of individual simulations with fixed insolations, in order to obtain a continuous progression of stratospheric humidity with insolation at an affordable computational cost. Although ExoCAM usually takes 40-50 yrs to equilibrate starting from an arbitrary initial condition, our transient simulations turn out to be a reasonably well approximation of a series of fixed insolation simulations (demonstrated in the results section), possibly because the insolation is turned up smoothly without abrupt jump. In addition to the above default high and low obliquity experiments, we also perform a series of mechanism suppression experiments. The model setups are described in the results section.

\section{Results}
\label{sec:results}
\subsection{Moister upper atmosphere under high obliquity}
\label{sec:moist-upper-atm}
\begin{figure*}[htp!]
  \centering
  \includegraphics[page=1,width=\textwidth]{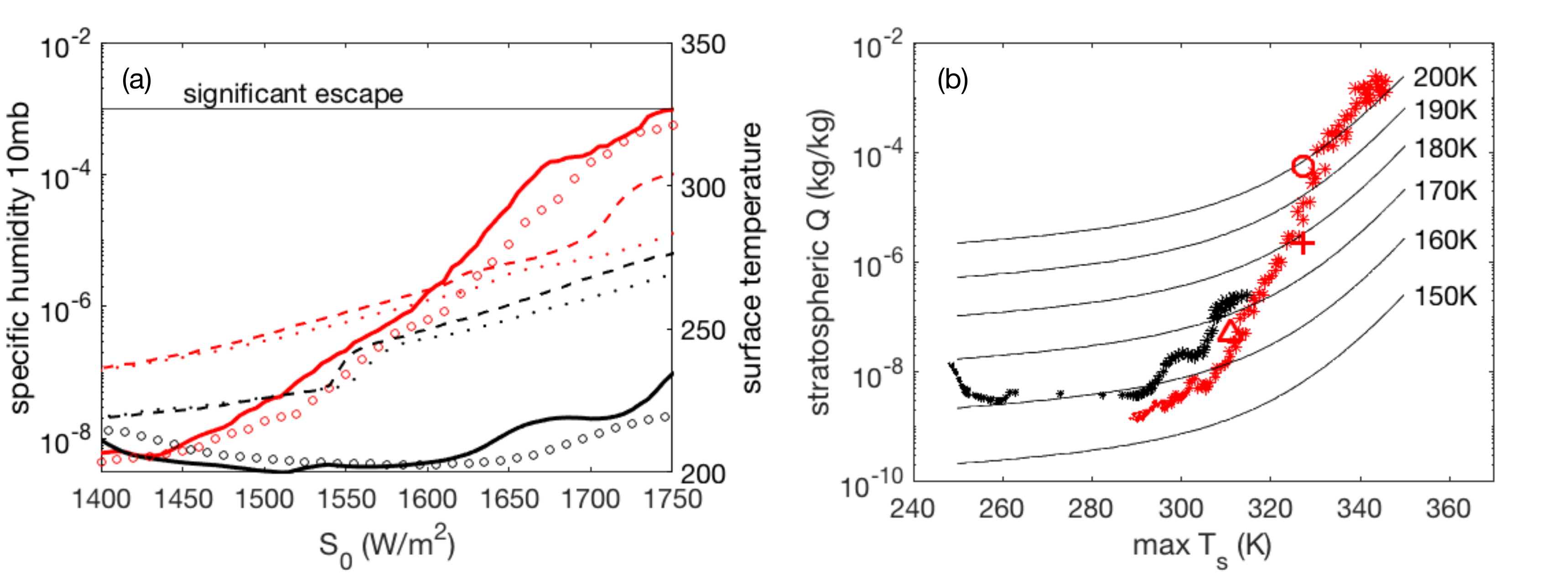}
  \caption{The evolution of upper atmospheric specific humidity and surface temperation with increasing insolation. Panel (a) shows the time series of the global mean specific humidity at 10 mb isobar in the solid curves (corresponding to the left axis), and shows that of the global annual mean surface temperature in the dashed curves (corresponding to the right axis). The high obliquity scenario is in red and the low obliquity is in black. The 1000 ppmv threshold for significant escape is marked by a thin black line. To demonstrate that the insolation change in the transient simulations is slow enough to allow climate to almost reach equilibrium, we repeat the simulation with insolation increased twice as fast. The progression of surface temperature (dots) and upper atmospheric specific humidity (circles) matches the slow-evolving transient experiment reasonably well. Panel (b) scatters the global-mean 10 mb specific humidity against the maximum monthly surface temperature achieved in that year (search among different latitudes and different months). High obliquity in red and low obliquity in black. Extra feedback suppression experiments are also marked in the plot. The red circle denotes a high obliquity experiment forced by fixed annual mean SST, the red ``+'' sign denotes a similar experiment except that the SST meridional distribution is reversed between the equator and the poles, and the red triangle denotes a high obliquity simulation forced by fixed annual mean insolation. Please refer to the text for more detailed model setups. For reference, the estimated water abundance in the upper atmosphere by 1D model is plot in thin black lines. We, following \citet{Kasting-Pollack-1983:loss}, assume moist adiabat from the surface until the temperature falls below the specified stratospheric temperature (marked to the right of each curve).}
  \label{fig:TS-Qstrat-line-scatter}
\end{figure*}

The upper atmosphere is significantly moister under high obliquity than low obliquity. The solid curves in Fig.~\ref{fig:TS-Qstrat-line-scatter}(a) show the progression of 10 mb\footnote{Choosing 1 mb gives almost identical results.} specific humidity, $Q_{10mb}$ with gradually increasing insolation for high and low obliquity scenarios. Starting from 1500 W/m$^2$, the upper atmospheric humidity surges up and reaches the 1000 ppm criteria for significant water escape around 1750 W/m$^2$. On the other hand, the upper atmosphere in the zero obliquity scenario remains very dry until 1750 W/m$^2$, which almost triggers the runaway greenhouse (adding another 50 W/m$^2$ insolation crashes the low obliquity model, but not the high obliquity one). This result has two implications. 1) Water vapor, which has evaporated from the surface, is more detectable under high obliquity. 2) High obliquity planets are more likely to become inhabitable due to water escape, while the key factor that determines the habitability on low obliquity planets tends to be the runaway greenhouse effect.

As shown in \citet{Kang-2019:mechanisms}, high obliquity planets tend to be tens of degrees warmer than their low obliquity equivalents. Shown in Fig.~\ref{fig:TS-Qstrat-line-scatter}(a) dashed curves are the progression of global annual mean surface temperature, $\overline{T}_s$, for the high and low obliquity scenarios. Despite the warmer climate, the upper atmosphere is still moister under high obliquity. For example, $\overline{T}_s=277K$ is achieved both at 1670 W/m$^2$ under high obliquity and at 1750 W/m$^2$ under low obliquity. The $Q_{10mb}$ corresponding to a 277K surface temperature is 2$\times$10$^{-4}$ kg/kg under high obliquity, over 3 orders of magnitude greater than that under low obliquity (1$\times$10$^{-7}$ kg/kg). At upper level, 4 mb, the specific humidity contrast between low and high obliquity experiments is even stronger, yielding 4 orders of magnitude difference, since the specific humidity drops more in the low obliquity experiment as moving upward. Therefore, the inadequacy of the 1D escape model \citep{Kasting-Pollack-1983:loss, Kasting-Whitmire-Reynolds-1993:habitable, Wordsworth-Pierrehumbert-2013:water} is clear -- with the same $\overline{T}_s$, 1D models will predict the same humidity in the upper atmosphere, however, they differ by 3-4 orders of magnitude when accounting for 3D circulation and seasonal variation.

\begin{figure*}[htp!]
  \centering
  \includegraphics[page=2,width=0.8\textwidth]{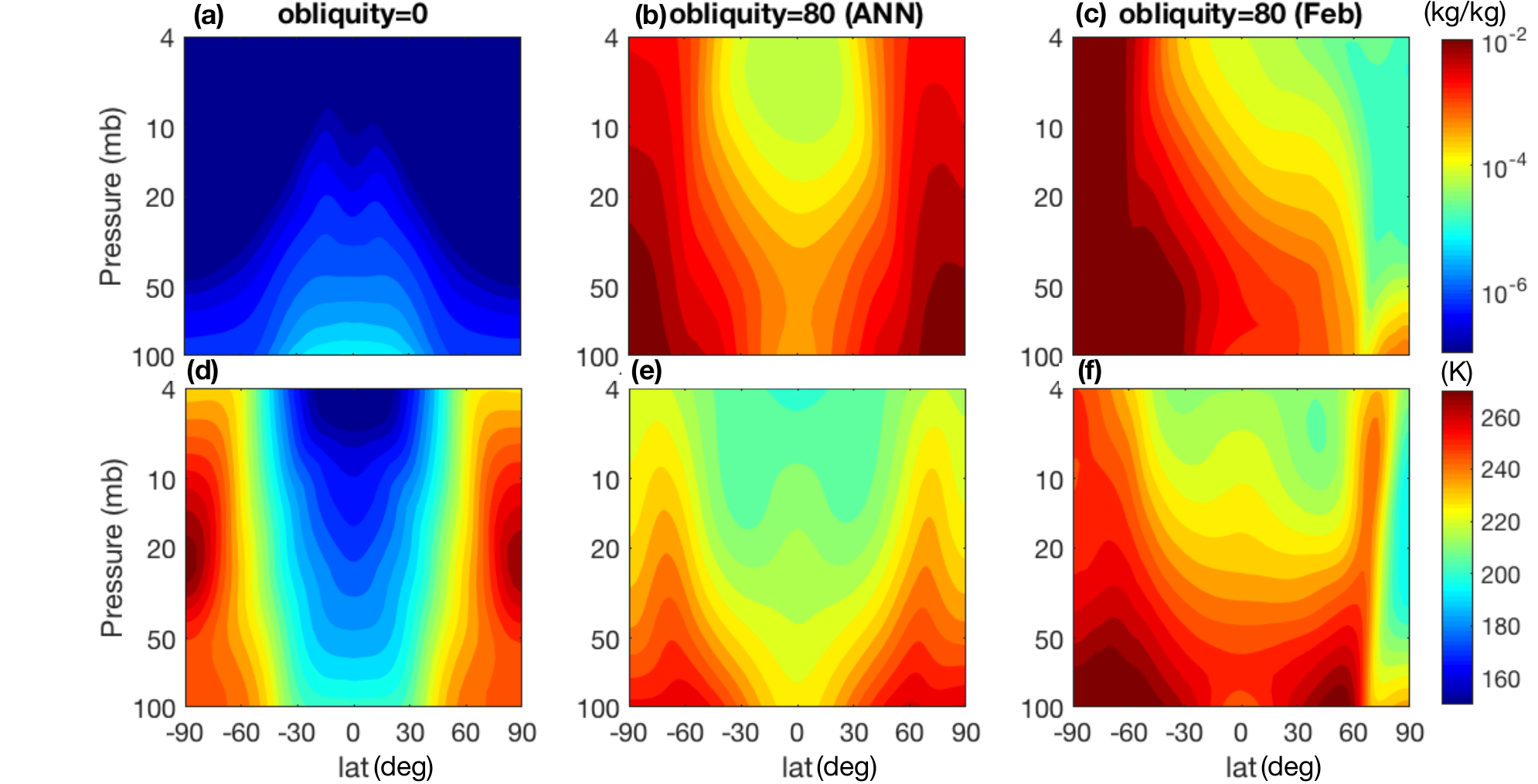}
  \caption{Meridional cross section of specific humidity (upper panels) and temperature (lower panels). Shown are 12-month averages in the transient simulations, centered at 1750 W/m$^2$ insolation forcing. (a,d) for zero obliquity, (b,e) for 80 obliquity annual mean, and (c,f) for 80 obliquity February. Climatology in simulations with fixed 1750 W/m$^2$ insolation resembles the approximation given by the transient simulations (not shown), indicating that the climate is fairly close to the equilibrium.}
  \label{fig:levlat-Q-T}
\end{figure*}

The spatial distributions of specific humidity and temperature are shown in Fig.~\ref{fig:levlat-Q-T}(a,d) for the zero obliquity experiment and Fig.~\ref{fig:levlat-Q-T}(b,e) for the high obliquity experiment, both at 1750 W/m$^2$ insolation. The zero obliquity experiment is close to the situation of the Earth. The air entering the stratosphere is primarily from the tropics, where the cold trap is located (Fig.~\ref{fig:levlat-Q-T}b). Therefore, this cold trap is able to freeze water vapor out as air parcels pass through it (see Fig.~\ref{fig:levlat-Q-T}a, the specific humidity decreases with altitude near the equator), filling the whole upper atmosphere with super dry air \citep{Holton-Haynes-McIntyre-et-al-1995:stratosphere, Butchart-2014:brewer}.

In contrast, the upper atmosphere under high obliquity is much moister (Fig.~\ref{fig:levlat-Q-T}b), because its troposphere is moister to begin with, particularly in the polar regions, and this moist air then enters the stratosphere\footnote{Upward motion occurs at both the subtropics and polar regions, but the subtropical air is cold and dry. This can be inferred from the $\Omega T_p$ term in Fig.~\ref{fig:T-budget}}, bypassing the cold trap at low latitudes (Fig.~\ref{fig:levlat-Q-T}e). The February climatology (Fig.~\ref{fig:levlat-Q-T}c,f) to a large extent resembles the annual mean (Fig.~\ref{fig:levlat-Q-T}b,e)\footnote{The difference between the two hemispheres gets much more significant in February, as the heat and moist stored from the previous summer gets exhausted (not shown).}, except that the summer hemisphere is slightly warmer and moister at high latitudes. 

Since air tends to enter the stratosphere from the hottest latitudes, we attempt to find a link between the stratospheric water vapor abundance and the maximum rather than the global annual mean surface temperature. In Fig.~\ref{fig:TS-Qstrat-line-scatter}(b), global average 10 mb specific humidity, $Q_{10mb}$, is scattered against the peak surface temperature achieved within a 12-month interval (high obliquity in red and zero obliquity in black). Within the brief overlap between the high and zero obliquity experiments, $Q_{10mb}$ in both experiments is roughly in the same order of magnitude, with the zero obliquity stratosphere being slightly moister. However, this does not mean that water enters the stratosphere under zero obliquity with more ease, since it is not in an appropriate comparison. In the zero obliquity experiment, the peak temperature always exists in the Equatorial region, which occupies a large portion of global surface area. However, the peak temperature only briefly shows up in the high obliquity experiment at the narrow polar regions during solstice, meaning that for most of the time, the stratospheric air parcels originate from a surface cooler than as marked in x-axis of Fig.~\ref{fig:TS-Qstrat-line-scatter}(b).

We, therefore, propose three hypotheses to explain the humid upper atmosphere under high obliquity. 1) The seasonal variation under high obliquity allows a temporary high surface temperature during polar days which would not occur without a seasonal cycle. 2) The cold trap loses efficiency under high obliquity since the moist air can enter the stratosphere without going through it. 3) The higher global annual mean surface temperature due to the low ice/cloud albedo under high obliquity increases the atmosphere's capacity to hold water in general \citep{Kang-2019:mechanisms}. In the rest of the letter, we will examine the three hypotheses through a chain of mechanism suppression experiments. 

\subsection{Mechanisms for the moist upper atmosphere under high obliquity.}
\label{sec:mechanism}

\begin{figure*}[htp!]
  \centering
  \includegraphics[page=3,width=\textwidth]{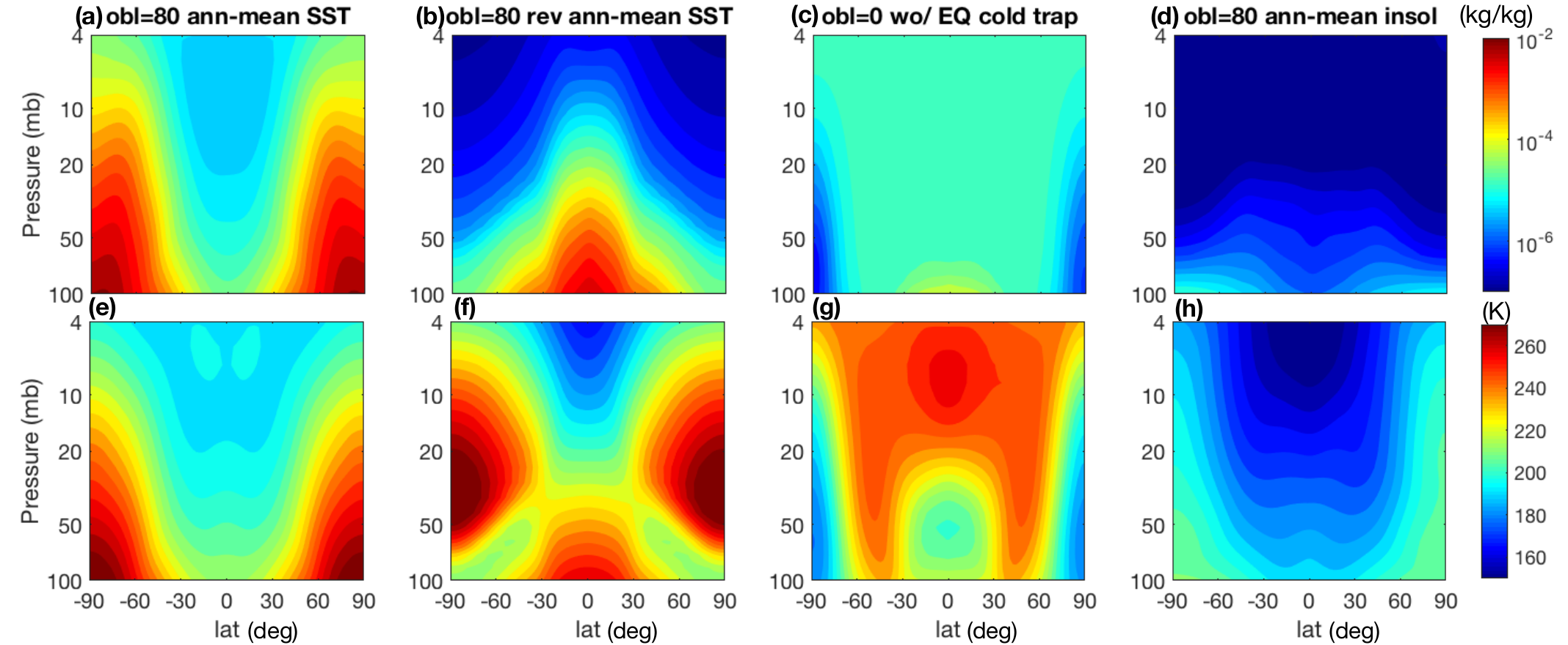}
  \caption{Same as Fig.~\ref{fig:levlat-Q-T}, but for the mechanism suppression experiments. (a,e) show the 80 obliquity experiment with fixed surface temperature which is the annual mean surface temperature in the default experiment. The experiment setups in (b,f) are the same as (a,e), except that the surface temperature distribution is flipped over about 45N/S. (c,g) show the fixed-SST 0 obliquity experiment without equatorial cold trap. (d,h) show the 80 obliquity experiment forced by annually averaged insolation.}
  \label{fig:levlat-Q-T-suppression}
\end{figure*}

To examine hypothesis 1), we first remove the seasonal variation of the sea surface temperature (SST), by fixing it to the annual-mean value from the high obliquity experiment with 1750 W/m$^2$ insolation. The zonal mean specific humidity and temperature are shown in Fig.~\ref{fig:levlat-Q-T-suppression}(a,e) respectively. Without the hot polar day time, the specific humidity in the upper atmosphere drops by 2 orders of magnitude, from 1000 to 10 ppmv, indicating that the seasonal variation does lead to a much moister stratosphere. This annual mean SST experiment is marked in Fig.~\ref{fig:TS-Qstrat-line-scatter} as an empty red circle, and it falls in the envelope of the other red dots, meaning that the decrease of upper atmospheric humidity is roughly consistent with the drop in peak SST. Even though the stratosphere gets 100 times drier without seasonal variation of SST, it is still 100 times moister than the low obliquity equivalents, which has only 0.01 ppmv water vapor at 10 mb on global average. This indicates that although the seasonal variation makes a difference, it is yet enough to explain the humidity difference between the low and high obliquity experiments.

\begin{figure*}[htp!]
  \centering
  \includegraphics[page=4,width=\textwidth]{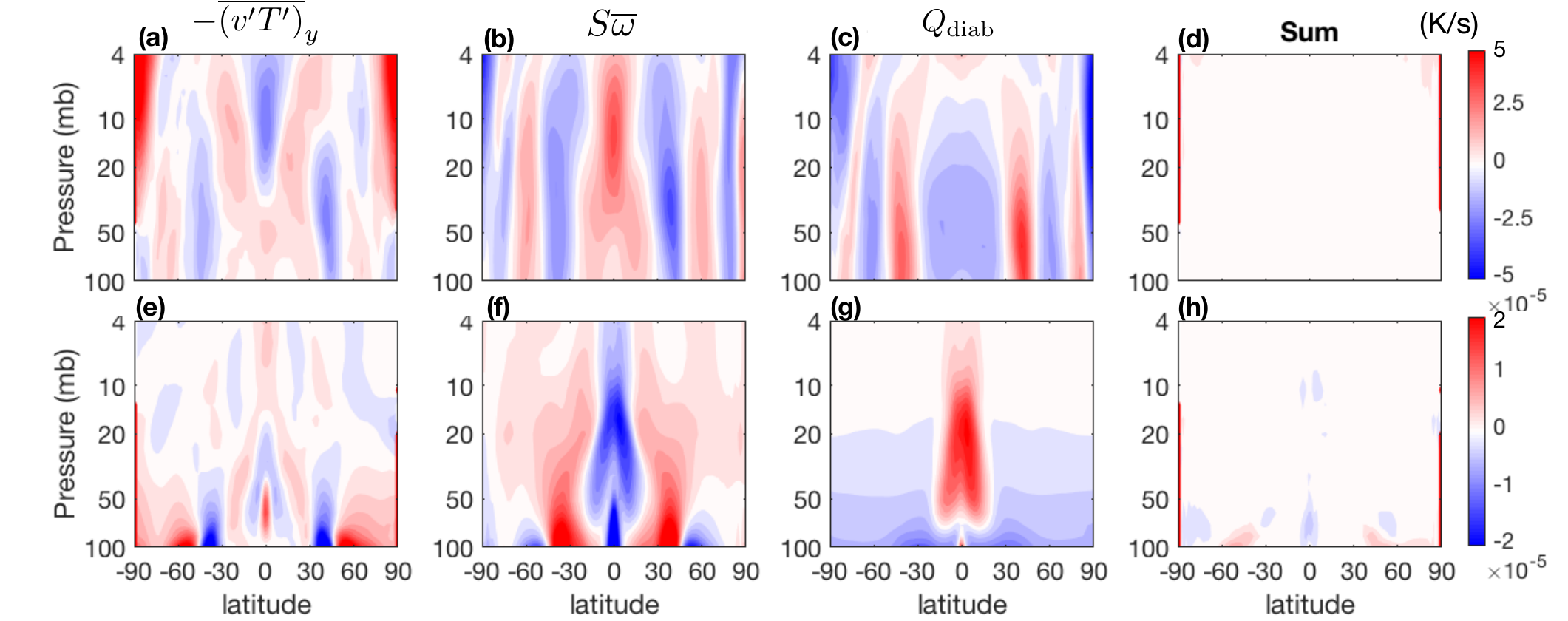}
  \caption{Temperature budget for (upper) the high obliquity and (lower) the low obliquity experiments under 1750 W/m$^2$ insolation. From left to right are the heating rate due to eddy meridional transport $-\overline{(v'T')}_y$, due to adiabatic heating $S\overline{\omega}$ ($S$ is static stability defined as $S=-\overline{T}\ln(\overline{\theta})_p$), and the diabatic heating $Q$, and sum of all budget terms. The vanishing of the ``sum'' suggests a closed budget. Other terms like $-\overline{(\omega'T')}_p$, $-\overline{V}\overline{T}_y$ are negligible compared to the terms shown here.}
  \label{fig:T-budget}
\end{figure*}

The hypothesis 2) is related to the relative location between the cold trap and the hottest SST. Under low obliquity situation, they overlap with each other over the Equator, as on the Earth \citep{Holton-Haynes-McIntyre-et-al-1995:stratosphere}. Upward motion above the Equator injects air into the stratosphere, and at the same time, causes adiabatic cooling, as demonstrated in Fig.~\ref{fig:T-budget}f. Under high obliquity, the poles are warmer than the equator as more radiation is received there, however, the cold trap remains located at low latitudes, as it is under low obliquity (Fig.~\ref{fig:levlat-Q-T}e,f). The reason behind this has been demonstrated in \citet{Faulk-Mitchell-Bordoni-2017:effects} and \citet{Singh-2019:limits}. The meridional movement of an air parcel is highly constrained at high latitudes, due to the strong meridional angular momentum gradient there. As a result, even with the substellar point at the pole, the strongest upward motion would not occur at the summer pole. Instead, according to \citet{Singh-2019:limits}, it occurs around the mid-latitudes, when the planet's rotation rate is close to the present-day Earth's. Shown in Fig.~\ref{fig:T-budget} upper panels are the dominant temperature budget terms for the high obliquity experiment. As expected, upward motion around 30-50N/S leads to adiabatic cooling there, and meridional eddy heat transport\footnote{could be simply due to the temporal correlation between the zonal mean $V$ and $T$ across different seasons} exports heat to the subtropics, forming a cold trap in the low latitudes. Although upward motion also occurs at the high latitudes (Fig.~\ref{fig:T-budget}b), the resultant cooling is counterbalanced by the heating induced by the poleward eddy heat transport $\overline{v'T'}$ (Fig.~\ref{fig:T-budget}a), which has been investigated by \citet{Kang-Cai-Tziperman-2019:tropical} using 1D and 3D simple models. Therefore, it seems that a cold trap will form above the low latitudes by the adiabatic cooling induced by upward motion there, regardless of the obliquity. This result holds in all experiments we show here, and the cold trap strength increases with rotation rate.

Given that the cold trap is offset from the highest SST under high obliquity, we expect water to enter the stratosphere more easily (hypothesis 2). To examine how much moistening is induced by this mechanism, we repeat the previous fix-SST experiment, except that we invert the SST meridional distribution about 45N/S. This way, the highest (lowest) SST is moved to the Equator (poles) with the absolute values unchanged. It is worth noticing that the global annual surface temperature ends up being increased, as the equatorial band has a larger area weight, and this could potentially increase the stratospheric humidity. However, as shown in Fig.~\ref{fig:levlat-Q-T-suppression}(b) and the ``+'' mark in Fig.~\ref{fig:TS-Qstrat-line-scatter}(b), the stratospheric humidity drops from 10 ppmv to 0.1 ppmv, with the SST meridional distribution being reversed. As a verification, we took the zero obliquity experiment and added an external heating source above the equator in the stratosphere to remove the cold trap there. The resultant stratospheric humidity increases from 0.01 ppmv to over 100 ppmv (Fig.~\ref{fig:levlat-Q-T-suppression}c). This suggests that the ineffectiveness of the cold trap under high obliquity could lead to an additional leap of stratospheric water abundance. 

In addition to the above two mechanisms, the fact that the climate is warmer under high obliquity \citep{Kang-2019:mechanisms} could also moisten the stratosphere. To examine this, we turn off the seasonal variation of the radiation completely by imposing the annual mean insolation at each latitude for all seasons\footnote{Note that the cold trap is still offset from the peak surface temperature meridionally and thus the mechanism proposed in hypothesis 2) still works.}. As a result, the global annual mean (peak) surface temperature drops from 306K (327K) to 292K (311K), which is even slightly lower than the zero obliquity equivalents \citep[for detail mechanisms, readers are referred to][]{Kang-2019:mechanisms}. The global mean stratospheric specific humidity thus drops significantly from 10 ppmv to 0.02 ppmv (red triangle in Fig.~\ref{fig:TS-Qstrat-line-scatter}b). The humidity spatial distribution (Fig.~\ref{fig:levlat-Q-T-suppression}d) clearly demonstrates a very dry stratosphere under annual mean insolation, and the difference between the high and zero obliquity experiment completely disappears. The stratosphere could be dried further (not shown), if this surface temperature distribution is reversed about 45N/S so that the cold trap would be functional.


\section{Conclusions}
\label{sec:conclusions}
We investigated the stratospheric humidity using a 3D general circulation model, under low obliquity and high obliquity conditions. Through a wide range of insolation, the high obliquity scenario was found to have much higher stratospheric water vapor abundance than its low obliquity equivalents. Even with the same global annual mean surface temperature, the stratospheric humidity under high obliquity could be 3-4 orders of magnitude greater than under low obliquity, something not captured by 1D escape models \citep{Kasting-Pollack-1983:loss, Kasting-Whitmire-Reynolds-1993:habitable, Wordsworth-Pierrehumbert-2013:water}. We then ran a series of mechanism suppression experiments to examine the role played by several complexities that exist only under high obliquity: 1) the seasonal variation of the surface temperature, 2) the ineffectiveness of the cold trap due to its meridional offset from the highest surface temperature, and 3) the warmer climate in general as shown in \citet{Kang-2019:mechanisms}. We found evidence showing that each of these complexities could contribute to the increased stratospheric humidity.

Our work focuses on only two parameters, the insolation and the obliquity, while fixing the others. Therefore, cautions need to be taken before generalizing our conclusions to an actual exoplanet. Rotation rate could make a difference. In the sensitivity test, we found the cold trap to become less and less evident as the self rotation slows down, meaning that the second mechanism might become negligible in the end of slow rotation.  Also, the relative importance of mechanism 1) and 3) is expected to change with surface heat inertia. Greater heat inertia would reduce the SST seasonal variation, which is necessary to mechanism 1), but it would warm up the high obliquity climate in general \citep{Kang-2019:mechanisms}, enhancing mechanism 3).
The atmospheric composition could not only change the vertical and horizontal distribution of temperature, but also give rise to extra sinks and sources for water vapor. In addition, we note that the real equilibrium climate can only be achieved when the parameter variation is infinitely slow, and thus the curve in Fig.~\ref{fig:TS-Qstrat-line-scatter}a can only be considered as an estimation to the equilibrium climate. Steady state simulations with fixed insolations are needed to give an accurate estimation of the inner edge of the habitable zone, which is not the main purpose of this study. 

For Earth-like planets, which have a rotation rate faster than several days, a liquid surface to provide enough heat inertia, an atmosphere that is primarily transparent to shortwave radiation (meaning that the atmosphere is heated from the surface), our results indicate: 1) we are more likely to be able to detect water evaporated from the surface in high obliquity planets; 2) with high insolation, high obliquity planets are more likely to lose habitability due to moist greenhouse effect, while the low obliquity planets are more likely to undergo runaway greenhouse.

\acknowledgments
The author thanks Prof. Eli Tziperman and Prof. Ming Cai for the insightful and helpful discussion. This work was supported by NASA Habitable Worlds program (grant FP062796-A) and NSF climate dynamics AGS-1622985. We would like to acknowledge high-performance computing support from Cheyenne provided by NCAR's Computational and Information Systems Laboratory, sponsored by the National Science Foundation. The relevant model outputs are archived in \href{https://www.dropbox.com/sh/hew5svg2hoi2bst/AABgFcNBCTiz2v7HWVZvUTi6a?dl=0}{here}.

\end{document}